# Ultra-low Energy, High-Performance Dynamic Resistive Threshold Logic

Mrigank Sharad, Deliang Fan and Kaushik Roy, *Fellow, IEEE*
School of Electrical and Computer Engineering, Purdue University, West Lafayette, Indiana 47907, USA
msharad@.purdue.edu, dfan@purdue.edu, kaushik@purdue.edu

*Abstract:* We propose dynamic resistive threshold-logic (DRTL) design based on non-volatile resistive memory. A threshold logic gate (TLG) performs summation of multiple inputs multiplied by a fixed set of weights and compares the sum with a threshold. DRTL employs resistive memory elements to implement the weights and the thresholds, while a compact dynamic CMOS latch is used for the comparison operation. The resulting DRTL gate acts as a low-power, configurable dynamic logic unit and can be used to build fully pipelined, high-performance programmable computing blocks. Multiple stages in such a DRTL design can be connected using energy-efficient low swing programmable interconnect networks based on resistive switches. Owing to memory-based compact logic and interconnect design and high-speed dynamic-pipelined operation, DRTL can achieve more than two orders of magnitude improvement in energy-delay product as compared to look-up table based CMOS FPGA.

## 1. Introduction

In recent years several computing schemes have been explored based on nano-scale programmable resistive elements, generally categorized under the term 'memristor' [1-3]. Of special interest are those which are amenable to integration with state of the art CMOS technology, like memristors based on Ag-Si filaments [2, 3] or spintronic memristors based on domain-wall magnets and magnetic tunnel junctions [9, 10]. Such devices can be integrated into metallic crossbars to obtain high density crossbar memory arrays. Some of these devices can facilitate the design of multi-level, non-volatile memory [3, 4].

The device-technologies for non-volatile resistive memory have led to possibilities of implementing programmable computing hardware that combine logic with memory. One such scheme is threshold logic [5-8]. It constitutes of summation of weighted inputs, followed by a threshold operation (fig. 1a), as given in eq.1 :

$Y = \text{sign}(\sum In_i W_i + b_i)$         (1)

Here, $In_i$, $W_i$ and $b_i$ are the inputs, weights and the thresholds respectively. While a memristor-array can be employed to perform current-mode analog summation of input signals, the thresholding operation requires the application of a comparator circuit. In recent proposal [5-7], such circuits have been designed using analog CMOS units that can be complex in terms of area and in-efficient in terms of power consumption. Such circuits based on analog amplifiers and current mirrors may suffer from stringent mismatch constraints and hence may not be scalable. Recently, design of resistive threshold logic gates (RTLG) based on Ag-Si memristor was demonstrated [8], where authors employed a simple CMOS flip-flop as a comparator. However, such a scheme would require application of large voltage and static current across the memristors in order to achieve enough sensing margin, leading to large power consumption.

In this work we propose the design of dynamic resistive threshold logic using programmable resistive elements. Noting that threshold logic synthesis with small fan-in TLGs require lower comparator resolution, we employ compact, low-power and high-speed dynamic CMOS latch for thresholding operation. Such hybrid dynamic RTLGs can be pipelined to achieve high performance, thereby leading to energy-efficient computation.

Rest of the paper is organized as follows. In section-2 we present the dynamic resistive threshold logic (DRTL) circuit. The design of DRTL array is described in section-3. Performance of the proposed scheme is discussed in section 4. A brief discussion on the choice of resistive memory elements is given in section-5. Section-6 concludes the paper.

## 2. Design of Dynamic Resistive Threshold logic Gate

For the design of DRTL-gate (DRTLG) we exploit the fact that, threshold logic synthesis using TLGs with small fan-in (2 to 3) need fewer levels in input weights as well as reduced comparator resolution (minimum % difference between threshold and input-summation to be detected) [11]. The set of weight levels needed for different fan-in restriction is depicted in fig. 1b, which shows that for a fan-in restriction of 2, only two weight-levels are required. The number of levels in the threshold was found to be 4 in this case. Lower number of weight-levels implies higher variation tolerance for weights (fig. 1b) and relaxed resolution constraint for the comparison operation. For instance for ideal 2-level weights, a 2-input TLG requires a comparator of only 25% resolution. The figure also shows that the increase in number of nodes while reducing the fan-in restriction from 4 to 2 is only marginal. Hence, owing to the aforementioned advantages offered by lower fan-in restrictions, in this work we limit our discussion on 2-input DRTLG.

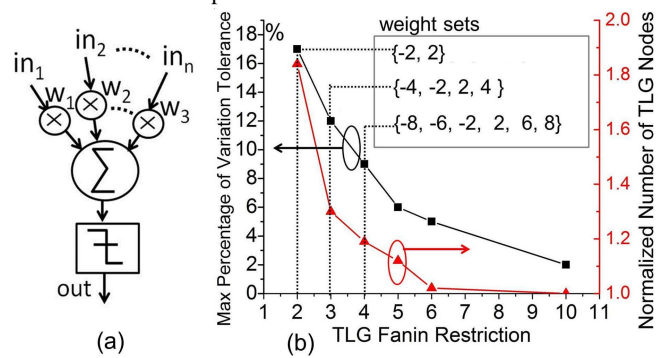

Fig. 1(a) A threshold logic gate, (b) Change in number of TLG required for a given logic block for different TLG restrictions.

A relaxed resolution constraint for thresholding operation can facilitate the use of compact and high speed dynamic CMOS latch for the comparison operation, required for the RTLG design. Note that the design of fixed, non-programmable threshold logic using such dynamic CMOS latches has been previously proposed in literature [12]. The programmable DRTLG unit proposed in this work is shown in fig. 2. It constitutes of cross-coupled CMOS inverters forming a regenerative feed-back based comparator circuit. It can detect the difference in the effective resistance in the two pull-down paths (connected to the source terminals of M3 and M4). The pull-down path constitutes of three branches corresponding to the two inputs and

the bias. The gates of the input transistors (M8 and M10) are driven by the two inputs of the

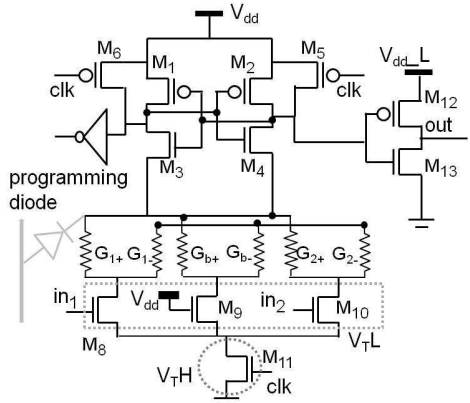

Fig. 2 Circuit for 2-fan in DRTL gate

DRTLG, whereas that of the bias transistor M9, is connected to $V_{dd}$. These transistors connect to the positive and the negative pull down paths of the latch using programmable conductance's, $G_{i+}$ and $G_{i-}$ respectively. Note that the sign of the weights, as well as, the threshold of a TLG can be either positive or negative. The resistive elements connected to the three transistors are programmed so as to effectively provide the corresponding weights. For instance, for M8, $G_+ - G_-$ should be proportional to the weight for input $in_1$ of the TLG and should have the same sign as $G_+ - G_-$. The range of the resistance values can be chosen high enough such that the impact of the *ON*-resistance of the transistors become insignificant in logic evaluation. This can allow the use of minimum size transistors for the latch, leading to a compact design [13].

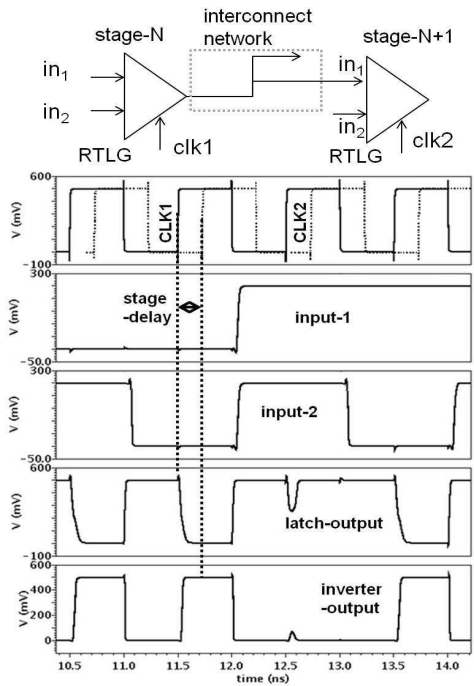

Fig. 3 Simulation plots for two input DRTLG operation

Fig. 3 depicts the simulation waveform for the DRTL gate shown in fig. 2. Before the inputs arrive, the clock input to the latch is low, which leads to pre-charging of the output nodes. After the inputs arrive, the clock can be applied leading the input dependent transition at the output nodes. The settled output of the latch is then available to its fan-out gates. Two-phase clocks, as shown in the plot, can be used for high performance. In this scheme, the clock period for logic evaluation is determined by the maximum delay in the latch evaluation. This delay can be less than 0.5 ns for 45nm CMOS technology. Note that, owing to the use of dynamic CMOS latch, the static current path in DRTL circuit is active for a small fraction of the total evaluation period, thereby leading to small static power consumption [13].

## 3. Design of pipelined DRTL array

Fig. 4 shows a fully pipelined threshold logic network. It consists of TLGs arranged in multiple pipelined logic stages. In order to facilitate field programmability for the proposed DRTL scheme, apart from the gate level configurability, the interconnects between the RTLG levels need to be programmable. This can be achieved by the use of resistive crossbar memory array as shown in fig. 5. The range of resistance values in the resistive memory for the interconnect design is critical to the overall energy efficiency. High *ON-OFF* ratio, with low on-resistance would be desirable for energy-efficient signaling and low off-state leakage. In this work an *ON* and *OFF* resistances of ~200Ω and 10MΩ were respectively chosen. Such resistance ranges may be achievable for CMOS compatible Ag-Si memristors [2, 3] ( a brief discussion on the device choice for DRTL circuit is given in section-5).

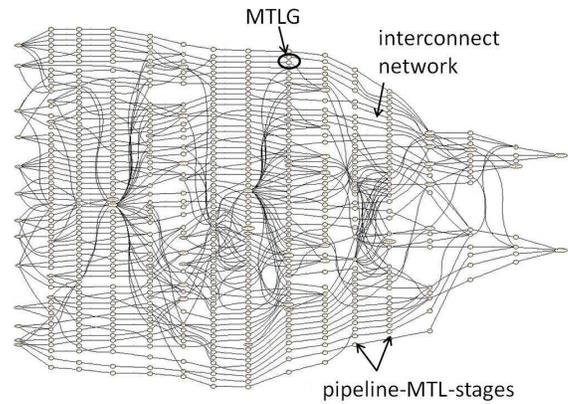

Fig. 4 Pipleined threshold logic network for the benchmark C-432

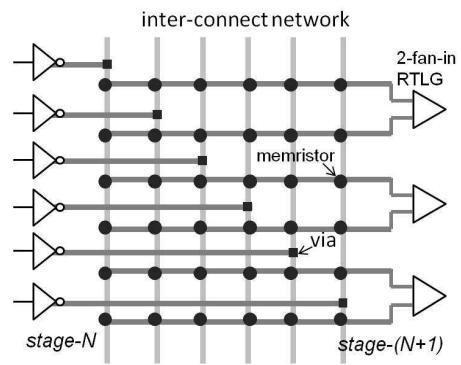

Fig. 5 Schmatic for interconnect design for DRTL using resistive crossbar memory

In this scheme each stage of a DRTL design can be mapped into multiple logic bloks with configurable interconnects. In order to minimize the energy dissipation in the interconnects, low-voltage signaling can be employed. This can be achieved by employing a low-supply-voltage for the output inverters for the DRTL gates as shown in fig. 2. Simultaneously, the input transistors can be chosen to have low $V_t$ in order to maintain low on-resistance even for reduced input voltage swing. The increase in leakge current in the DRTL circuit due the use of low $V_t$ input transistros can be mitigated by the using high-$V_t$ transistors for clocking as shown in fig. 2. In this work we chose a sclaed $V_t$ of ~130mV for the input transistors, wheras the nominal $V_t$ for 45nm transistors was ~350mV. This allowed the use of ~250mV for interconnects while the $V_{dd}$ used for the DRTL circuit was 0.5V.

## 4. Design Performance

Based on the foregoing techniques for DRTLG and interconnect design, we evaluated the performance of the proposed logic scheme. Table-1 compares the performance of DRTL with 4-input LUT based CMOS FPGA [11], for some ISCAS-85 benchmarks. As mentioned before, the dual functionality of computing and memory provided by the programmable resistive elements can be potentially exploited for energy efficient programmable logic and interconnect design. DRTL exploits such nano-devices to realize fully dynamic and high-performance computing blocks. The dynamic operation of the CMOS latches minimizes static-power dissipation, generally dominant in resistive logic schemes proposed in literature [5-8]. The pipeline scheme on the other hand provides the maximum possible throughput for a given gate-level performance. Average energy dissipation per DRTL-gate was found to be less than ~0.3 fJ, while that for interconnect-driving was ~0.2fJ (per fan-out, based on crossbar layout). This led to less than ~1fJ per DRTL gate on an average for 2GHz throughput. As compared to CMOS FPGA, results show more than ~96% lower energy dissipation for most benchmarks, while more than 100X lower energy-delay product can be possibly achieved.

**Table-1 Performance Comparison: DRTL vs. CMOS-LUT**

| ISCAS-85 benchmark | # input | # output | delay /throughput (ns) | | Energy (fJ) | | % reduction | |
|---|---|---|---|---|---|---|---|---|
| | | | LUT | RTL | LUT | RTL | energy | energy-delay |
| c432 | 36 | 7 | 10.1 | 0.5 | 17362.56 | 480 | 97.2 | 99.86 |
| c499 | 41 | 32 | 8.18 | 0.5 | 33795.57 | 940 | 94.5 | 99.83 |
| c880 | 60 | 26 | 8.4 | 0.5 | 26394.41 | 970 | 96.3 | 99.78 |
| c1355 | 41 | 32 | 9.95 | 0.5 | 56284.24 | 1480 | 97.4 | 99.87 |
| c1908 | 33 | 25 | 11.55 | 0.5 | 56930.13 | 1200 | 97.89 | 99.91 |

## 5. Choice of Resistive Memory

As mentioned before, high-*ON-OFF* ratio is essential for minimizing the energy dissipation in the interconnects, which can be provided by CMOS compatible memristors. For the interconnect devices, precise values of resistance levels may not necessary, as long as the *ON* and *OFF* values are within a certain limit. Thus, high-speed bi-state switching can be employed for the memristors employed in the interconnect design, for changing the state of the resistive-devices between the two extremes. Owing to the dynamic operation, the *ON-OFF* ratio of the resistive elements in the DRTL gates does not affect the energy-efficiency significantly. Hence, magnetic memory can be employed to implement the computational weights in the DRTL gates. The ratio of the resistance levels in binary magnetic spin-torque memory can be limited to a factor of ~3 to 4. However, the resistance of the two binary states are fixed and are determined by device parameters like the thickness of the tunneling oxide used in the magnetic tunnel junctions (MTJ). This can facilitate fast and energy efficient programming. On the other hand, obtaining sufficiently accurate resistance levels with Ag-Si memristors may require analog-write techniques [3]. Fig. 6 shows 2-terminal and three terminal, multi- multilevel magnetic memory cells suitable for DRTL design [10].

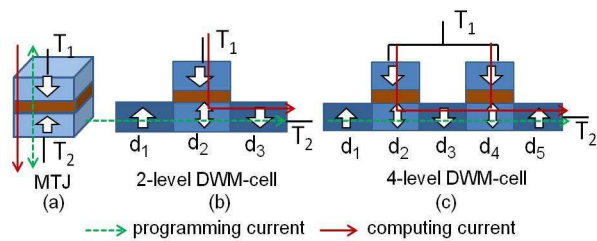

Fig. 6 (a) 2-terminal MTJ, (b) 3-terminal, two level domain wall switch, where d2 is the 'free' switchable domain (c) three terminal, 4-level domain wall switch with d2 and d4 as free domains [10].

Programming for both spin and Ag-Si memristor based resistive memory can be done with the help of diodes or dedicated access transistors, the prior being more area area-efficient method [14].

## 6. Conclusion

In this work we proposed the design of Dynamic Resistive Threshold Logic (DRTL) that employs non-volatile resistive memory elements for reconfigurable computing. DRTL is a dynamic, pipelined logic scheme that can achieve low power consumption and high performance. Comparison with 4-input LUT-based CMOS FPGA shows the possibility of ~96% higher energy efficiency and more than two orders of magnitude lower energy-delay product for DRTL.

**Acknowledgment:**

This work was funded in parts by SRC and CSPIN